\documentclass[prl,twocolumn,reprint,tightenlines,nopacs,showkeys,superscriptaddress,groupedaddress]{revtex4} 
\usepackage{graphicx} 
\usepackage{dcolumn} 
\usepackage{bm} 
\usepackage{amssymb} 
\usepackage{comment} \usepackage{subfigure} \usepackage{amsmath}
\usepackage{float} \usepackage{natbib} \usepackage{tabularx}
\usepackage{ dsfont } \usepackage{cancel} \usepackage{pdfpages}

\newcommand*\xbar[1]{%
  \hbox{%
    \vbox{%
      \hrule height 0.5pt 
      \kern0.5ex
      \hbox{%
        \kern-0.1em
        \ensuremath{#1}%
        \kern-0.1em
      }%
    }%
  }%
} 

\begin{document}

\title{Extrinsic and intrinsic correlations in molecular information
  transmission}

\author{Vijay Singh} \affiliation{Department of Physics, Emory
  University, Atlanta, GA 30322, USA}\affiliation{Computational Neuroscience Initiative, University of
  Pennsylvania, Philadelphia, PA 19104, USA}
\author{Martin  Tchernookov} \affiliation{Department of Physics, Emory
  University, Atlanta, GA 30322, USA}\affiliation{Department of Physics, Lamar University, Beaumont, TX 77710}
\author{Ilya  Nemenman} \affiliation{Department of Physics, Emory
  University, Atlanta, GA 30322, USA}
\affiliation{Department of Biology, Emory
  University, Atlanta, GA 30322, USA}

\begin{abstract}
  Cells measure concentrations of external ligands by capturing ligand
  molecules with cell surface receptors. The numbers of molecules
  captured by different receptors co-vary because they depend on the
  same extrinsic ligand fluctuations. However, these numbers also
  counter-vary due to the intrinsic stochasticity of chemical
  processes because a single molecule randomly captured by a receptor
  cannot be captured by another. Such structure of receptor
  correlations is generally believed to lead to an increase in
  information about the external signal compared to the case of
  independent receptors. We analyse a solvable model of two molecular
  receptors and show that, contrary to this widespread expectation,
  the correlations have a small and {\em negative} effect on the
  information about the ligand concentration. Further, we show that
  measurements that average over multiple receptors are almost as
  informative as those that track the states of every individual one.
\end{abstract}

\maketitle

{\em Introduction.}  Information processing is a crucial function of
life \cite{tkacik:2016}.  It typically involves representing external
signals by activities of biological elements, such as cell receptors,
genes, or neurons. A lot is known about information processing by such
individual elements \cite{Bialek:1991ub, Ziv:2007bo,Tkacik:2008us,
  Tkacik:2008dq,Tostevin:2009uk, Tkacik:2011jr,Cheong:2011jp,
  Nemenman:2012tb,Fairhall:2012hk}. However, the fascinating phenomena
emerging in information processing by many interacting biological
elements are only beginning to be uncovered \cite{Averbeck:2006ew,
  Walczak:2010tj,Tkacik:2010hk,
  daSilveira:2013vf,Hormoz:2013kw,Hu:2014cy,mugler2016limits,tkacik:2016}.

A particularly well-developed example of multivariate biological
information processing is population coding by neurons
\cite{ginzburg1994theory,sompolinsky2001population,
  sompolinsky1999effect,abbott1999effect,pola2003exact,
  Averbeck:2006ew,Hu:2014cy,da2014high,shamir2014emerging,moreno2014information}.
   Here many neurons (often heterogeneous
and interacting) are treated as conveying information about the same
stimulus. A celebrated general property of such networks is the ``sign
rule'' \cite{Averbeck:2006ew,Hu:2014cy}, which suggests that if
fluctuations of neural activities due to changes in the signal are
orthogonal to fluctuations due to intrinsic coupling among the
neurons, then the collective of neurons has more information about the
stimulus than a collective of noninteracting neurons would have.

Deriving the sign rule requires making serious (though often implicit)
assumptions about the structure of fluctuations in populations of
sensors. Verifying these assumptions is hard for networks as complex
as those in the brain. In contrast, multiple receptors on the cell
surface are a cellular biology equivalent of population coding in
neuroscience, with an advantage that the structure of correlations
among the sensors (receptors) does not have to be postulated {\em a
  priori}, but can be derived analytically from biophysically
plausible molecular interactions.  We use this advantage to study
collective information processing in an analytically solvable model of
two receptors interacting via binding to the same chemical ligand
species. We show, in particular, that the sign rule is violated in
this system, and the information gathered about the stimulus by the
interacting receptors is smaller than in the noninteracting case. This
suggests that studies of population codes based on correlations are
insufficient (including in computational neuroscience, where they are
common) since effects of the correlations depend on features of
biophysical mechanisms that establish them.

In addition to its illumination of the limitations of the {\em
  general} sign rule, the two receptors model addresses an important
question {\em specific} to cellular information processing.
Estimation of a chemical signal concentration by cells has been
studied since the seminal work of Berg and Purcell \cite{Berg:1977bp},
with notable new recent results
\cite{bray1998receptor,Bialek2005,Endres:2008eb,endres2009maximum,hu2010physical,kaizu2014berg,mugler2016limits,singh2015}. However,
most of these formulations consider the combined (or averaged)
response of all receptors on the cell surface for estimating the
concentration. Keeping track of responses of individual receptors
would provide extra information about the concentration stored in the
receptor-to-receptor variability.  Our model quantifies how useful it
is for the cell to keep track of such data. We show that, for large
observation times, the average population response is almost as
informative about the stimulus as the set of activities of all
individual receptors.

{\em Background.}  We introduce the sign rule with the following
simple yet instructive model \cite{Averbeck:2006ew,Hu:2014cy}. Imagine
a Gaussian signal $s$ with the mean $\bar{s}$ and the variance
$\sigma^2_s$. It is measured by two responses, $r_1$ and $r_2$ (firing
rates of neurons or receptor activity). For simplicity, these are
assumed linearly and equivalently dependent on $s$ (or the response to
small fluctuations is linearized), such that
\begin{equation}
  r_1=as+\eta_1,\quad
                   r_2=as +\eta_2,\label{example2}
\end{equation} 
where $a$ is the gain, and $\eta_{1,2}$ are Gaussian noises with
$\langle \eta_i\rangle=0$, and ${\rm var}\, \eta_i \equiv \langle
\eta_i^2 \rangle=\sigma_\eta^2$. 

We estimate the signal from the responses as
$s_{\rm est}= (r_1+r_2)/(2a)$. Then the estimation error variance is 
\begin{equation} 
{\rm var}\, (s_{\rm est}-s)\equiv \sigma^2_{\rm err}=
\frac{\sigma_\eta^2(1+\rho_\eta)}{2a^2}.
\label{simpleerror}
\end{equation}
Here $\rho_\eta \sigma^2_\eta={\rm cov}\, (\eta_1,\eta_2)$ stands for
the covariance of the two noises, or the {\em noise-induced}
covariance \cite{Averbeck:2006ew}, and $\rho_\eta$ is the
corresponding correlation coefficient. By analogy with the intrinsic
noise in systems biology \cite{Swain:2002kn}, $\rho_\eta$ can also be
called the {\em intrinsic noise correlation}. When $\rho_\eta=0$,
Eq.~(\ref{simpleerror}) reduces to the usual decrease of the error
variance by a factor of two for two independent measurements. However,
when $\rho_\eta<0$, the error variance is smaller. In particular, if
$\rho_\eta\to-1$, the signal can be estimated with no
error. Generalizing this simple observation, one can define the {\em
  stimulus-induced response covariance} \cite{Averbeck:2006ew} or the
{\em extrinsic noise covariance} \cite{Swain:2002kn}, as the
covariance between mean responses to stimuli, averaged over all
stimuli,
$ {\rm cov}\, (\bar{r}_1,\bar{r}_2)\equiv \rho_sa^2\sigma_s^2$. Then
our example illustrates the {\em sign rule} \cite{Hu:2014cy}: if
$\rho_s$ and $\rho_\eta$ are of opposite signs, then the stimulus can
be inferred from the two responses with a smaller error compared to
the (conditionally) independent responses, $\rho_\eta=0$. The same
result can restated using {\em mutual information} between the two
responses and the stimulus
\cite{Shannon:1998ti,Nemenman:2012tb,tkacik:2016,SI} :
\begin{equation}
  I[{r_1,r_2};s]=\frac{1}{2} \ln \left[ 1+ \frac{a^2
      \sigma_s^2}{(1+\rho_\eta) \sigma^2_\eta}\right].
\label{Inf_rho}
\end{equation}
For Eq.~(\ref{example2}), $\rho_s=1>0$, and then $\rho_\eta<0$
corresponds to increase in the information.

In the case of a chemical ligand being absorbed by two identical
receptors, the mean values of $r_1$ and $r_2$ change in the same way
with the ligand concentration, so that $\rho_{\rm s}=1>0$. At the same
time, a molecule absorbed at one receptor cannot be absorbed at the
other, which should give $\rho_\eta<0$, and hence will increase the
measured information according to the sign-rule. However, in
computational neuroscience, where these ideas originated, noise
(co)-variances are inferred empirically and are, in principle,
unconstrained. In contrast, in cell biology, intrinsic noises are
generated from the discreteness and stochasticity of individual
chemical reaction events
\cite{Elowitz:2002hb,Paulsson:2005uz,VanKampen:2011vs}, which
constrains relations among these quantities. In particular,
$\rho_\eta$ may depend on $\sigma_\eta$, and then it is unclear if the
sign rule would hold in Eq.~(\ref{Inf_rho}). Indeed, the primary
contribution of this Letter is to show that measuring the ligand
concentration with two identical receptors does not obey the sign
rule.

\begin{figure}
\begin{center}
\includegraphics[scale=0.2]{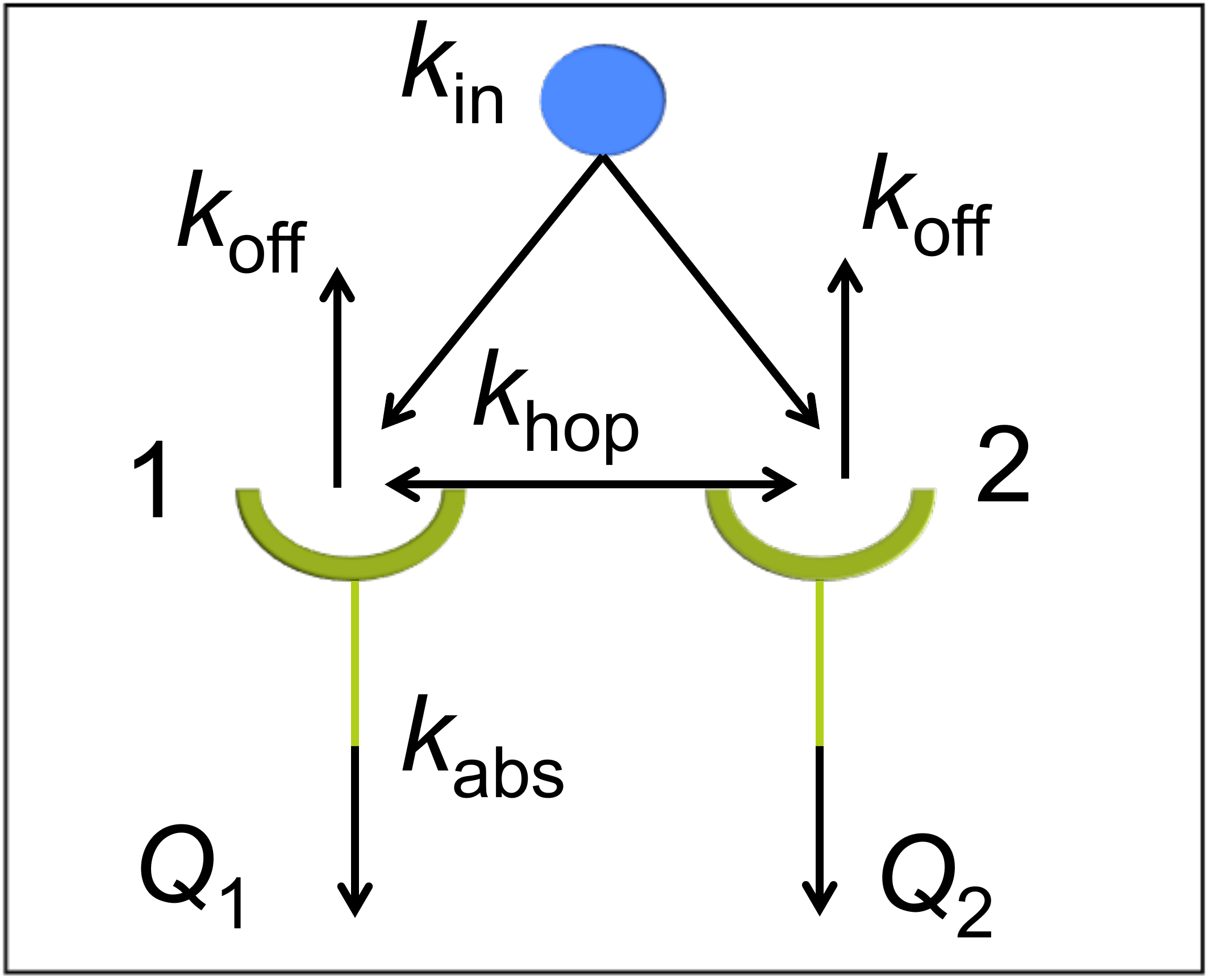}
\caption{Model schematics. Receptors 1 and 2 bind ligands with rate
  $k_{\rm{in}}$, and the bound molecules can detach and diffuse away
  to infinity with the rate $k_{\rm off}$. The bound ligands also can
  be absorbed with the rate $k_{\rm{abs}}$, or they can dissociate and
  diffuse to the other receptor (hop) with the rate
  $k_{\rm{hop}}$. $Q_{i}$ is the number of ligands absorbed at the
  receptor $i$. \label{two_receptor}}
\end{center}
\end{figure}

{\em The Model}.  We consider two identical receptors that can bind
ligand molecules with a rate $k_{\rm{in}}$
(Fig.~\ref{two_receptor}). No more than one molecule can be bound to
each receptor at the same time (with no restrictions on the number of
bound molecules, the dynamics is linear, the receptors are
conditionally independent). The bound molecule can be
absorbed/deactivated with the rate $k_{\rm{abs}}$, freeing the
receptor (absorbing receptors collect more information about the
stimulus compared to binding-unbinding receptors
\cite{Endres:2008eb}). Alternatively, it can unbind and leave the
vicinity of receptors with the rate $k_{\rm off}$. Finally, it can
leave one receptor and diffuse to the other. We model this as a hop
between the receptors with the rate $k_{\rm{hop}}$, which in reality
would depend on the diffusion constant and the distance between the
receptors.  The number of molecules absorbed on both receptors over
time $t$, $\{Q_1(t),Q_2(t)\}$, carries information about the binding
rate $k_{\rm in}$. Since, $k_{\rm in}$ is proportional to the ligand
concentration, such counting of the absorbed molecules measures the
concentration.

Within this setup, we investigate how the ligand-induced interaction
between the two receptors affects the information about the
concentration, $I [Q_1,Q_2;k_{\rm in}]$, cf.~Eq.~(\ref{Inf_rho}). Note
that the hopping can change the conditional distribution
$P(Q_1,Q_2|k_{\rm in})$, which can affect the information, but it
cannot change the conditional distribution of the total number of
captured molecules $Q_+=Q_1+Q_2$.  Thus the change in the information,
if any, can come only from the dependence between $Q_-=Q_1-Q_2$ and
$k_{\rm in}$. This expands the molecular sensing literature
\cite{Berg:1977bp,Bialek2005,Endres:2008eb}, where one typically
estimates $k_{\rm in}$ based only on the integrated number of observed
ligands, $Q_+$. In other words, together with our main question, we
will quantify if the set of individual responses of all receptors,
$\{Q_1,Q_2\}$ or $\{Q_+,Q_-\}$, is more informative about the
concentration than the integrated response alone.

{\em Solution.}  To calculate the distribution
$P(Q_1,Q_2|k_{\rm{in}})$, we start with the master equation describing
the dynamics of the vector of probabilities of having 0 or 1 molecules
bound to each of the receptors,
${\mathbf P}=\{P_{ij};\,
i,j=0,1\}^T=\{P_{00},P_{01},P_{10},P_{11}\}^T$,
\begin{equation}
  \dot {\mathbf P}(t)=-H\,{\mathbf P}(t).
\end{equation}
Here the generator matrix is
\begin{equation}H=
\begin{bmatrix}
2 k_{\rm{in}} & -k_{\rm{off}}- k_{\rm{abs}} &  -k_{\rm{off}} - k_{\rm{abs}} & 0 \\ 
-k_{\rm{in}}  & k_{\rm{tot}} & -k_{\rm{hop}} & - k_{\rm{abs}}-k_{\rm{off}}\\ 
-k_{\rm{in}}  & -k_{\rm{hop}} & k_{\rm{tot}} & - k_{\rm{abs}}-k_{\rm{off}}\\ 
0 & -k_{\rm{in}}  &-k_{\rm{in}}  &  2 k_{\rm{off}} + 2 k_{\rm{abs}}
\end{bmatrix},
\end{equation} with $k_{\rm{tot}} = k_{\rm{in}} +k_{\rm{off}}+k_{\rm{abs}}+k_{\rm{hop}}$.

To find the probability distribution of $Q_1$ and $Q_2$, we use the
generating functional technique
\cite{Bagrets:2003bc,Jordan:2004we,Gopich:2006jr,Sinitsyn:2007dt,Sinitsyn:2007wo,SI}.
Namely, we separate out the parts of $H$ that correspond to the
absorption events
\begin{align}
  H&\equiv H_0+H_{{\rm abs},1}+H_{{\rm abs},2},\\
H_{{\rm abs},1}&=
\begin{bmatrix}
0& - k_{\rm{abs}} & 0 & 0 \\ 
0 &0 &0 & 0\\ 
0  & 0 &  0& - k_{\rm{abs}}\\ 
0 &0  &0 &  0
\end{bmatrix},\\
H_{{\rm abs},2}&=
\begin{bmatrix}
0& 0&- k_{\rm{abs}}  & 0 \\ 
0 &0 &0 &  - k_{\rm{abs}}\\ 
0  & 0 &  0&0\\ 
0 &0  &0 &  0
\end{bmatrix}.
\end{align}
Then we tag the terms corresponding to the absorption reactions by
counting fields $e^{\chi_1}$ and $e^{\chi_2}$, forming the tagged
generator matrix,
\begin{align}
  \tilde{H}(\chi_1,\chi_2)&\equiv H_0+H_{{\rm
      abs},1}e^{\chi_1}+H_{{\rm abs},2}e^{\chi_2}.
\end{align}
Finally we realize that the vector of moment generating functions (or
the Laplace transforms) of $P(Q_1,Q_2|k_{\rm in},i,j)$, denoted as
${\mathbf
  Z}(\chi_1,\chi_2,t)=\left\{Z_{00},Z_{01},Z_{10},Z_{11}\right\}$,
satisfies the tagged master equation
\begin{align}
\dot {\mathbf Z}(\chi_1,\chi_2,t)=-\tilde{H}(\chi_1,\chi_2){\mathbf
  Z}(\chi_1,\chi_2,t).\label{modified_master}
\end{align}

We are interested in the long-time asymptotic, where each receptor has
had many absorption events, $Q_1,Q_2\gg 1$. Then the solution of
Eq.~(\ref{modified_master}) can be approximated as
\begin{align}
{\mathbf
  Z}(\chi_1,\chi_2,t)\approx {\mathbf
  Z}(0)\exp[{-\tilde{\lambda}_{\min}}(\chi_1,\chi_2)\,t],
\end{align}
where $\tilde{\lambda}_{\min}$ is the smallest real part eigenvalue of
$\tilde{H}$.  From here, one can read off the cumulant generating
functions conditional on the occupancy of the receptors, to the
leading order in $t$,
${
  F}_{ij}(\chi_1,\chi_2,t)\approx-\tilde{\lambda}_{\rm{min}}(\chi_1,\chi_2)\,t$.
As expected, the leading order is the same for any value of
$i,j$. Thus the means and the (co)variances of the numbers of absorbed
molecules, conditional on $k_{\rm in}$ all scale linearly with
time. They can be obtained by differentiating
$\tilde{\lambda}_{\rm{min}}(\chi_1,\chi_2)$ with respect to $\chi_1$
and $\chi_2$. Denoting by $\left<\dots|k_{\rm in}\right>$ expectations
conditional on $k_{\rm in}$, we write:
\begin{align}
  \left<Q_i|k_{\rm in}\right>&=t\left. \frac{\partial \tilde{\lambda}_{\rm
        min}(\chi_1,\chi_2,t)}{\partial
      \chi_{i}}\right|_{\chi_1,\chi_2=0},
  \\
  \left<\delta Q_{i}\delta Q_j|k_{\rm in}\right>&=t \left.\frac{\partial^2 { \tilde{\lambda}}_{\rm
        min}(\chi_1,\chi_2,t)}{\partial \chi_i\partial
      \chi_j}\right|_{\chi_1,\chi_2=0}.
\end{align}
In its turn, the eigenvalue $\tilde{\lambda}_{\rm min}$ can be
obtained using non-Hermitian perturbation theory considering $\chi_i$
as the perturbation parameters around the eignevalue
$\lambda_{\rm{min}}=0$ of the unperturbed Hamiltonian \cite{SI}. For compactness
of notation, we define
$k_{\rm ioa}=k_{\rm in}+k_{\rm off}+k_{\rm abs}$. This gives:
\begin{align}
  \left<Q_i|k_{\rm in}\right>&=\frac{k_{\rm{in}}
    k_{\rm{abs}}\;t}{k_{\rm ioa}}, 
\end{align}
\begin{align}
  \left<\delta Q_i\delta Q_i|k_{\rm in}\right>&=  \left<Q_i|k_{\rm
      in}\right>\nonumber\\
&\hspace{-10mm}\times\left(1-\frac{2
    k_{\rm{in}} k_{\rm{abs}}}{k_{\rm
      ioa}^2}
  + \frac{2k_{\rm{hop}}k_{\rm{in}}k_{\rm{abs}}}{k_{\rm
      ioa}^2(k_{\rm{tot}}+k_{\rm{hop}})}\right),
\label{sig_nn} \\
\left<\delta Q_1\delta Q_2|k_{\rm in}\right>&=- 2\left<Q_i|k_{\rm
      in}\right>\frac{k_{\rm{hop}} k_{\rm{in}}
  k_{\rm{abs}}}{k_{\rm ioa}^2(k_{\rm{tot}}+k_{\rm{hop}})}. 
\label{sig_nm}
\end{align}
These expressions fully define the conditional distribution
$P(Q_1,Q_2 | k_{\rm{in}})$ to the leading, Gaussian order. Notice that
$ \left<\delta Q_1\delta Q_2|k_{\rm in}\right><0$ as long as
$k_{\rm hop}\neq0$, and thus, according to the sign rule, we expect
more information from the two correlated receptors than the two
independent ones with $k_{\rm hop}=0$.

In the basis of
$Q_{\pm}=Q_1\pm Q_2$, the covariance matrix diagonalizes, and we get
\begin{align}
  &\left<Q_+|k_{\rm in}\right>=\frac{2\;k_{\rm{in}} k_{\rm{abs}}\;t}{k_{\rm ioa}}, \\
  &\left<Q_-|k_{\rm in}\right>=0, \\
  &\left<\delta Q_+^2|k_{\rm in}\right>=\left<Q_+|k_{\rm in}\right>\frac{[k_{\rm ioa}^2- 2
    k_{\rm{in}} k_{\rm{abs}}]}{k_{\rm ioa}^2},  \\
  &\left<\delta Q_-^2|k_{\rm in}\right>=\left<Q_+|k_{\rm in}\right>\frac{k_{\rm ioa}^2- 2
      k_{\rm{in}} k_{\rm{abs}}+2k_{\rm hop}k_{\rm ioa}}{k_{\rm ioa}(k_{\rm{tot}}+k_{\rm{hop}})}, \\
&\left<\delta Q_+\delta Q_-|k_{\rm in}\right>=0.
\end{align}
Since neither $ \left<Q_+|k_{\rm in}\right>$ nor
$ \left<\delta Q_+^2|k_{\rm in}\right>$ depend on $k_{\rm hop}$, these
expressions clearly show that the total number of molecules absorbed
by the two receptors is not affected by the interaction parameter
$k_{\rm{hop}}$, as we alluded to previously. The coupling between the
receptors only affects the variance of the difference of the number of
molecules coming from each receptor.

We now define the absorption currents $J_{\pm}=Q_{\pm}/t$, so that
$ \left<J_\pm|k_{\rm in}\right>= \left<Q_\pm|k_{\rm in}\right>/t$, and
$\left<\delta J^2_\pm|k_{\rm in}\right>=\left<\delta Q^2_\pm|k_{\rm
    in}\right>/t^2$.
Now assuming a Gaussian marginal distribution of $k_{\rm in}$, with
the mean $\bar{k}_{\rm in}$ and the variance $\sigma^2_{k_{\rm in}}$,
we write down the marginal distribution of absorption currents
averaged over the external signal concentrations
\begin{multline}
  P(J_+,J_-)=\int \frac{dk_{\rm{in}}}{\sqrt{2\pi}\sigma_{k_{\rm
        in}}}\exp\left[{-\frac{(k_{\rm in}-\bar{k}_{\rm
          in})^2}{2\sigma^2_{k_{\rm in}}}}\right]\\
\times\frac{\exp \left[-\frac{(J_+-\left<J_+|k_{\rm in}\right>)^2}
{2  \left<\delta J_+^2|k_{\rm in}\right>}-\frac{J_-^2}{2
     \left<\delta J_-^2|k_{\rm in}\right>} \right]}
{2\pi
   \sqrt{\left<\delta J_+^2|k_{\rm in}\right>\left<\delta J_-^2|k_{\rm in}\right>}}.
\label{full_joint_dist}
\end{multline}
Note that $\left<\delta J_\pm^2|k_{\rm in}\right>\propto 1/t$ for
large $t$. This is the usual manifestation of the law of large
numbers, so that the ratio of the standard deviation of the currents
to their means decreases as $\propto 1/t^{1/2}$.

Both $\left<J_+|k_{\rm in}\right>$ and
$\left<\delta J_\pm^2|k_{\rm in}\right>$ depend on $k_{\rm in}$. We
assume that $\sigma^2_{k_{\rm in}}$ is small, so that this dependences
can be written to the first order in
$\delta k_{\rm in }=k_{\rm in}-{\bar{k}_{\rm in}}$. Then the
dependence of the mean currents on $k_{\rm in}$ preserves the Gaussian
form of Eq.~(\ref{full_joint_dist}), while the dependence of the
variance manifests itself in sub-Gaussian orders. To the leading order
in small $\sigma^2_{k_{\rm in}}$, the marginal distribution of the
currents is still a product of two Gaussians,
\begin{align}
&P(J_+,J_-)=\frac{1}{2\pi\sigma_+\sigma_-}e^{-\frac{(J_+-
  \langle J_+\rangle)^2}{2\sigma_+^2}- \frac{J_-^2}{2\sigma_-^2}},\;
  {\rm with}\\
  &\left<J_+\right> = \frac{2\;\bar{k}_{\rm in}k_{\rm abs}}{\bar{k}_{\rm ioa}},\\
 & \sigma_+^2 =\left<\delta J_+^2|\bar{k}_{\rm
               in}\right>\left[1+ \left( \frac{\partial
               \left<J_+\right>}{\partial k_{\rm{in}}} \right)^2_{\bar{k}}
               \frac{\sigma_{k_{\rm in}}^2 }{\left<\delta J_+^2|\bar{k}_{\rm in}\right>
               }\right],\\
  &\sigma_-^2 =\left<\delta J_-^2|\bar{k}_{\rm in}\right>.
\label{joint_dist}
\end{align}

The mutual information we are seeking is
$I[Q_1,Q_2;k_{\rm{in}}]=S[Q_1,Q_2]-\left<S[Q_1,Q_2|k_{\rm{in}}]\right>_{k_{\rm{in}}}$,
where $S$ are the marginal and the conditional entropies.  In the
limit of small $\sigma_k^2$, entropies are given by logarithms of the
corresponding variances, so that
\begin{align}
I[Q_1,Q_2;k_{\rm{in}}]= \frac{1}{2} \ln \left[1+ \left(
  \frac{\partial \left<J_+\right>}{\partial k_{\rm{in}}} \right)^2
  _{\bar{k}} \frac{\sigma_k^2 }{\left<\delta J_+^2|\bar{k}_{\rm
  in}\right>  } \right],
\label{info_final}
\end{align}
which is independent of $k_{\rm{hop}}$.

The mutual information in Eq.~(\ref{info_final}) is independent of the
interaction between the receptors, {\em violating} the sign rule. The
reason for the violation is easy to trace: although the intrinsic
receptor correlations are negative, the quantity
$(1+\rho_\eta)\sigma_\eta^2=\left<\delta J_+^2|k_{\rm in}\right>$ in
Eq.~(\ref{Inf_rho}) is independent of $k_{\rm{hop}}$! The biophysics of
the problem conspires to ensure that the variance of the number of the
absorbed ligands on the individual receptors increases by exactly the
amount to counteract the receptor correlations to the Gaussian order
in fluctuations. The effect of the correlations can only be seen in
the higher order corrections. This answers our main question about the
generality of the sign rule.  Further, we note that the information in
Eq.~(\ref{info_final}) is independent of $J_-$. This answers the
second question: to the Gaussian order and for large $t$, keeping
track of differences between the individual receptors does not change
the amount of available information.

\begin{figure}
\begin{center}
\includegraphics[scale=0.27]{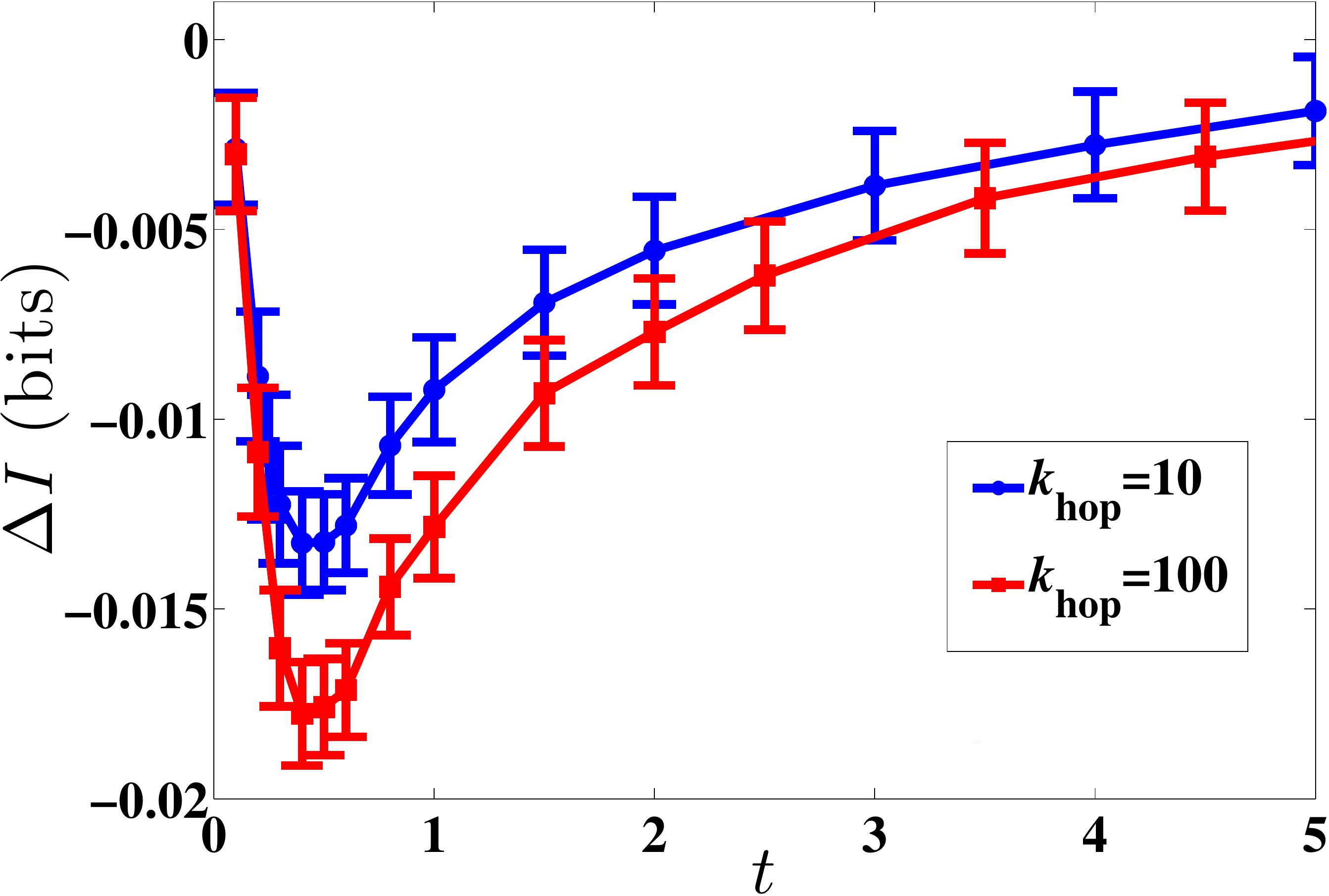}
\caption{Correlations due to molecular hopping reduce information
  about the signal. We plot the reduction in the information compared
  to two non-interacting receptors for
  $k_{\rm{in}}=k_{\rm{abs}}=10$. We use Gillespie
  \cite{Gillespie:2007bx} algorithm for simulations and NSB entropy
  estimator \cite{nemenman2001entropy} to evaluate information from
  data. Each point is obtained from $10^6$ samples from the steady state
  of the system dynamics for 17 values of $k_{\rm in}$.}
\label{simulation}
\end{center}
\end{figure}

To study non-Gaussian effects of hopping we evaluate
$\Delta I (\bar{k}_{\rm{in}},k_{\rm{abs}},k_{\rm{hop}}) =
I_{\bar{k}_{\rm{in}},k_{\rm{abs}},k_{\rm{hop}}}[Q_1,Q_2;k_{\rm in}] -
I_{\bar{k}_{\rm{in}},k_{\rm{abs}},0}[Q_1,Q_2;k_{\rm in}]$,
where the second term is equivalent to two independent receptors. We
simulate the system using the Gillespie algorithm
\cite{Gillespie:2007bx}. As illustrated in Fig.~\ref{simulation},
$\Delta I<0$, so that the receptor coupling through hopping {\em
  reduces} the mutual information, contradicting the very sign of the
sign rule. This is because the hopping introduces another stochastic
process into the system, increasing the overall noise. Further, at
$t\to\infty$, $\Delta I\to0$ for all hopping rates, indicating that
the receptor coupling does not provide extra information at large $t$
compared to independent receptors even to non-Gaussian orders.

{\it Discussion.} We have analyzed a simple model of two identical
receptors that are coupled through interactions with the same
ligand. Our main finding is that, in this system, the variance and the
co-variance of the receptor activities both depend on the interactions
between the receptors in such a way that the interactions do not
affect the amount of information between the receptor activities and
the ligand concentration to the Gaussian order in fluctuations. We
additionally discovered that the interactions have a {\em negative}
effect on the amount of available information in sub-Gaussian orders,
though the effect disappears at long observation time. These
observations violate the well-known ``sign rule''
\cite{Averbeck:2006ew,Hu:2014cy}. In contrast, in most previous
analyses, the variances of the individual sensors have been {\em
  assumed} independent of the interactions between the sensors
\citep{Averbeck:2006ew,ginzburg1994theory,seung1993simple,
  sompolinsky1999effect, abbott1999effect}, leading to the sign
rule. We show that biophysical interactions do not necessarily obey
such assumptions. We expect that similar concerns will be valid beyond
receptors in individual cells, in applications such as neural
population coding or multicellular molecular communication
\cite{Taillefumier:2015ky,mugler2016limits}. Thus such mechanistic
considerations must enter analyses of multivariate information
processing.

In studies of cellular sensing, one often make an assumption that
cells are only affected by the population-averaged activities of their
receptors. In principle, additional information about the external
ligand can be encoded in differences of activities of individual
receptors since these differences depend on the concentrations,
$Q_1-Q_2\sim\sqrt{k_{\rm in}}$. Our analysis provides a solid basis
for this assumption by showing that, for long observation times, the
cell has as much information about the signal when it tracks the sum
of activities of its receptors as if it were to track activities of
every individual receptor.

{\bf Acknowledgements.} This work was supported in part by James S.\
McDonnell Foundation grant 220020321 and NSF grants IOS-1208126 and
PoLS-1410978.

\bibliography{receptors}

\newpage
\widetext
\includegraphics[bb=1.0in 1.0in 7.5in 10in,page=1]{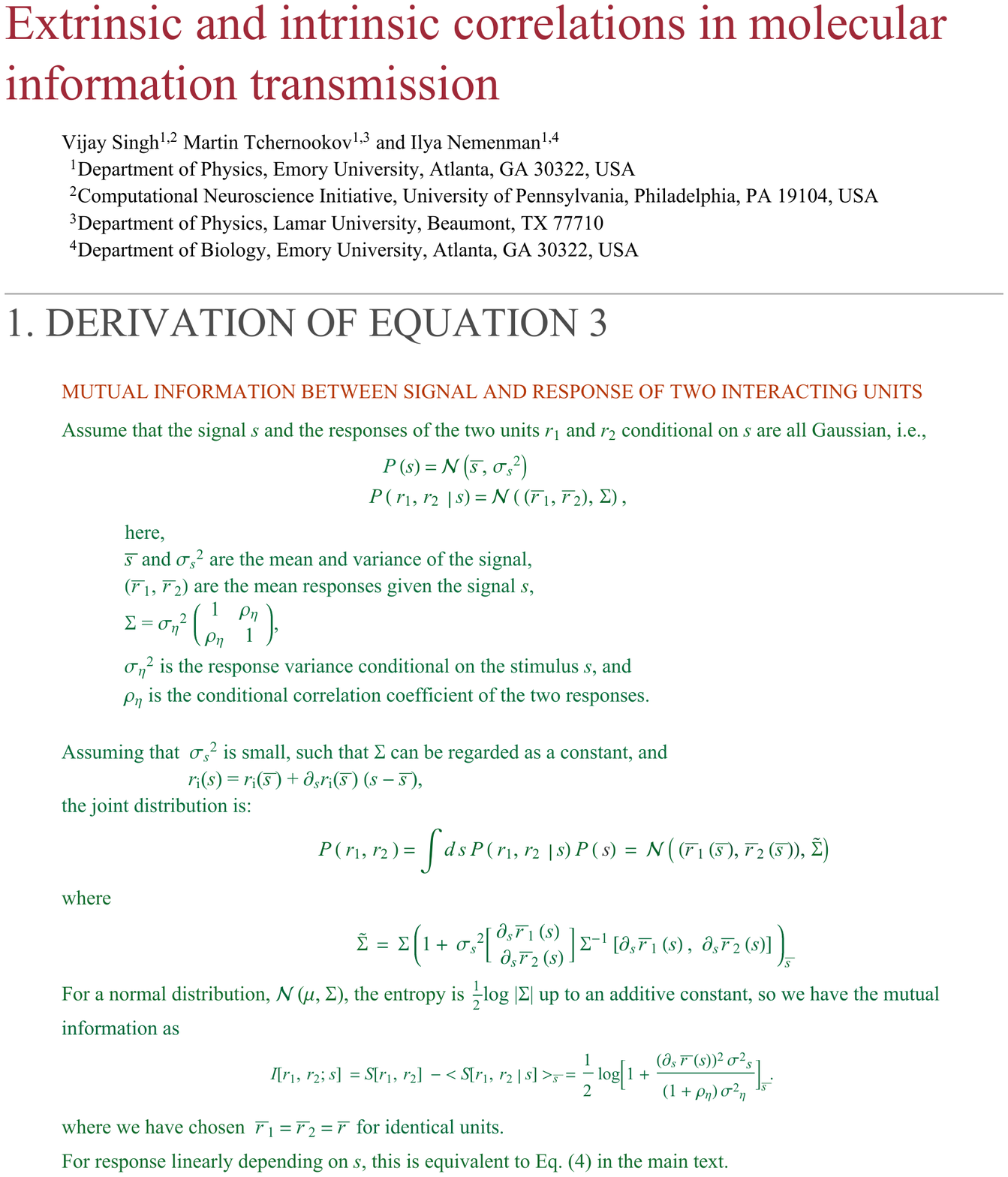}
\includegraphics[bb=1.0in 1.0in 7.5in 10in,page=2]{SI_Mathematica.pdf}
\includegraphics[bb=1.0in 1.0in 7.5in 10in,page=3]{SI_Mathematica.pdf}
\includegraphics[bb=1.0in 1.0in 7.5in 10in,page=4]{SI_Mathematica.pdf}
\includegraphics[bb=1.0in 1.0in 7.5in 10in,page=5]{SI_Mathematica.pdf}

\end{document}